\documentclass[twocolumn,aps,a4paper,prl,numerical]{revtex4-1}
\usepackage{amsmath}
\usepackage{epsf}
\usepackage{graphicx}
\usepackage{natbib}
\usepackage{times}

\begin{document}

\title{Clock-controlled emission of single-electron wavepackets in a solid-state circuit}
\author{
J.~D.~Fletcher,$^{1}$
M.~Kataoka,$^{1}$
H.~Howe,$^{2}$
M.~Pepper,$^{2}$
P.~See,$^{1}$
S.~P.~Giblin,$^{1}$
J.~P.~Griffiths,$^{3}$
G.~A.~C.~Jones,$^{3}$
I.~Farrer,$^{3}$
D.~A.~Ritchie,$^{3}$
T.~J.~B.~M.~Janssen, $^{1}$
}

\affiliation{$^{1}$ National Physical Laboratory, Hampton Road, Teddington, Middlesex TW11 0LW, United Kingdom}
\affiliation{$^{2}$ London Centre for Nanotechnology, and Department of Electronic \& Electrical Engineering, University College London, Torrington Place, London, WC1E 7JE}
\affiliation{$^{3}$ Cavendish Laboratory, University of Cambridge, J. J. Thomson Avenue, Cambridge CB3 0HE, United Kingdom }

\begin{abstract}
We demonstrate the transmission of single electron wavepackets from a clock-controlled source through an empty high-energy edge channel. The quantum dot source is loaded with single electrons which are then emitted with high kinetic energy ($\sim$150~meV). We find at high magnetic field that these electron can be transported over several microns without inelastic electron-electron or electron-phonon scattering. Using a time-resolved spectroscopic technique, we measure the electron energy and wavepacket size at picosecond time scales. We also show how our technique can be used to switch individual electrons into different paths.
\end{abstract}

\date{\today}

\maketitle

\begin{figure}
\includegraphics[width=8.6cm]{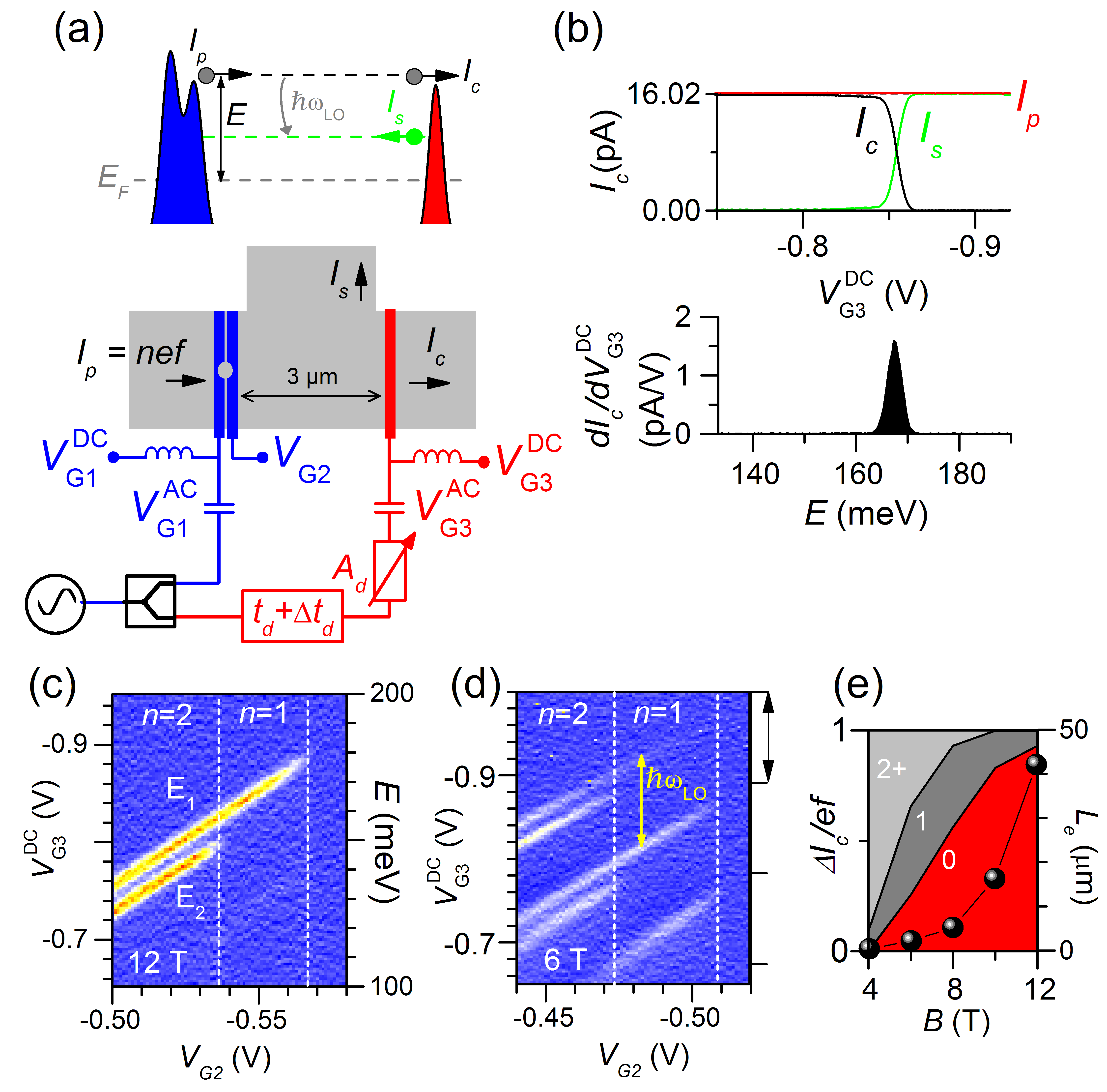}
\caption{%
(a)~Upper panel shows potential profile along the axis of the device. Electrons are ejected from the two-gate, tunable-barrier single-electron pump (blue) per cycle.  These are either transmitted through or reflected by the detector barrier (red). This is repeated at a frequency $f$, giving the measured pump, collector and side channel currents, $I_p, I_c$ and $I_s$. Lower panel shows a top view of the device with electrical connections. An RF signal generator operates both pump and detector barrier, the latter via a time delay, $t_d + \Delta t_d$ and attenuation $A_d$. %
(b)~Upper panel: $I_c$, $I_s$ and $I_p$ as the energy barrier is swept upwards, blocking the collector current. Lower: $dI_c/dV_{\rm G3}^{\rm DC}$ on calibrated energy scale for this detector (see supplemental information).
(c)~Color map of $dI_c/dV_{\rm G3}^{\rm DC}$ showing the position of ballistic peaks (bright lines) as a function of $V_{\rm G2}$ and $V_{\rm G3}^{\rm DC}$ in regions where $n=1, 2$ electrons are pumped per cycle.
(d)~As 1c but at a lower magnetic field where the emission of a phonon (pictured in 1a) gives side peaks in the electron energy distribution. %
(e)~Electron-phonon scattering effects as a function of magnetic field. Shaded areas correspond to the proportion of electrons undergoing 0,1, or more than 2 phonon emission events. Large circles show the scattering length $L_e$ calculated from this data.
}
\label{f1}
\end{figure}

A key goal in solid-state physics is to achieve coherent control of scalable quantum systems~\cite{Wallraff2004,You2002,Loss1998}. One strategy is to mimic devices used in the field of quantum optics, substituting optical sources, mirrors, and detectors with analogues based on single electrons in solid-state circuits \cite{Henny1999,Oliver1999,Bocquillon2012}. Low-dimensional electronic structures in magnetic fields should allow the coherent ballistic transmission of electrons between such components. While this has been observed in continuous sources formed through single-mode point contacts, revealing quantum effects with a true on-demand single-electron source requires controlling and measuring of the release time, electron energy and wavepacket size~\cite{LeichtSST2011,BattistaPRB2012}, as well as control of scattering mechanisms and sources of decoherence. Single-electron devices have been under development for some years for quantum device technology applications, including metrology~\cite{Pothier1992,Shilton1996,Keller1999} and quantum information processing~\cite{Loss1998,Petta2005,Koppens2006}.  An important part of this development involves driving single electrons into `single electron circuits' and demonstrating quantum mechanical effects, such as many-particle interference or entanglement~\cite{SamuelssonPRL2004,Oliver1999,Neder2007}. When injected into a conducting channel, these electrons have to be isolated from their environment, including interactions with other electrons and phonons.  Inelastic events will have destructive effects on the ballistic transport and coherence. Energy selectivity is required in order that electrons from the source can be separated from other background electronic excitations~\cite{Bocquillon2012}. Synchronizing the arrival of single electrons or measuring/controlling the wavepacket size is also extremely important, because timing variations may reduce the visibility of quantum effects even when the electrons fully maintain coherence. However, achieving full control over fast single-electron dynamics at picosecond timescales is extremely challenging~\cite{Shaner2002}.

Here, we demonstrate that, using a semiconductor dynamic quantum-dot system, we can attain fine control over the timing and size of single-electron wavepacket emission, with good suppression of inelastic scattering mechanisms.
In our devices, single-electrons are emitted at an energy of over 150~meV above the Fermi level at intervals controlled by a clock signal. Synchronous modulation of a detector barrier lying in the path of the electron beam allows us to measure the wavepacket size, which is found to be $\sim$ 80~ps. We also demonstrate that we can trap multiple electrons in the same quantum dot, emit these sequentially, and then use the small ($\sim$ 350~ps) separation in arrival time to split them into two different paths. This new method of manipulating a clock-controlled electron beam is a step towards fully controllable fermionic quantum-optics experiments~\cite{Hong1987,Einstein1935,Ansmann2009,Jonckheere2012,Giovannetti2006}.

Our single-electron source, whose operating principle~\cite{Kaestner2008,Kashcheyevs2010,Fletcher2012,Kataoka2011} and charge-transfer precision~\cite{Giblin2010,Giblin2012} have been studied extensively, is formed in a GaAs/GaAlAs two-dimensional electron gas (2DEG)~\cite{Note1} system by two surface gates, as shown in Fig.~\ref{f1}(a). An alternating voltage applied to one of the metal gates scoops electrons into a trap (a quantum dot), then ejects them, one at a time, out into a broad channel. Because the quantum dot is formed above the Fermi energy, the ejected electrons acquire an excess kinetic energy. In a large perpendicular magnetic field these electrons follow the edge of this channel in skipping orbits. In this experiment, after a distance of three microns, they encounter another gate-defined barrier. Depending on the barrier height with respect to the electron energy~\cite{Heiblum1985, Taubert2011}, this either reflects the electron into a side channel producing a current $I_s$, or allows it to pass into a collector channel giving current $I_c$. The barrier height is determined by the control voltages~\cite{Note2} applied directly to the barrier (see Fig. 1(a)), along with a small cross-talk contribution originating from electron pump itself (discussed later). The current in each output channel is measured with ammeters, which measure the probability of transmission/reflection, averaged over many cycles of pump operation.

First, we demonstrate the high-energy emission from the single electron source and our ability to control their emission energy. Figure~1(b) shows single-electron spectroscopy under a magnetic field $B = 12$~T perpendicular to the plane of the 2DEG and at an experimental temperature $T = 300$~mK. Pumping at a frequency of 0.1 GHz, a current $I_p \simeq ef \simeq 16.02$~pA of hot electrons is fired towards the barrier. As the detector barrier voltage $V_{\rm G3}^{\rm DC}$ is swept to more negative values, there is a sharp threshold beyond which this current is no longer guided into the collector channel ($I_c \simeq ef$ and $I_s = 0$) but is instead reflected into the side channel ($I_c = 0$ and $I_s \simeq ef$). This shows that the distribution of the hot electrons is almost entirely concentrated into a single, narrow ballistic peak~\cite{Heiblum1985, Palevski1989}, as seen in $d{I_c}/dV_{\rm G3}^{\rm DC}$ (Fig.~1(b)). In principle, the width of the energy distribution could be limited by several factors, including tunnel-broadening of the dot level, movement of the dot energy during emission or inelastic scattering (see supplemental information). The energy selectivity of the detector barrier will also limit the apparent peak width. The fact that these processes don't broaden the peak significantly shows that these processes are quite weak, particulary that there is very little inelastic scattering under these conditions.

By establishing the calibration between barrier height and gate voltage (see supplemental information) we find that the energy of the electrons emitted by the pump with respect to the Fermi energy is $E \sim 150~$meV. This is far larger than both the Fermi energy ($\sim 10~$~meV) and the temperature scale $k_BT <0.05$ meV. For certain quantum-optics type experiments, this is ideal as the large energy gives a clear separation between pumped electrons and thermal excitations~\cite{Bocquillon2012}.

Using this spectroscopic technique, we find that the emission energy is controlled by the height of the pump exit barrier, which determines how high the trapped electron potential has to be raised before electrons leave the pump. Figure~1(c) shows that this energy (labelled $E_1$) can be tuned linearly by varying the exit barrier gate voltage $V_{\rm G2}$, giving an \textit{in situ} control of the emission energy~\cite{LeichtSST2011}. The $E_2$ peak in Fig.~1(c) appears when pumping two electrons per cycle and is discussed later. This energy control could be used to match or de-tune the emission energy of devices in a single electron circuit.

For the ballistic measurements in Fig.~1(c) large magnetic fields were required to suppress inelastic processes. This strongly reduces electron-electron interactions, as described elsewhere~\cite{Taubert2011} and also influences electron-phonon scattering rates. At lower magnetic fields, as in Fig.~1(d), we see evidence for emission of one or more optical phonons by each electron. Optical phonon emission is the strongest electron-phonon process in GaAs~\cite{Sivan1989}. Because of the well-defined phonon energy, we observe that this divides the transition in $I_c$ into stages, repeating `phonon replicas' of the main spectral line spaced by the energy of the LO optical phonon $\hbar\omega_{LO} = $ 36~meV~\cite{Taubert2011}. We find that high magnetic fields ($B = 12$~T) massively reduces phonon emission, giving an estimated scattering length of $\gtrsim$ 40$\mu$m (see Fig.~1(e)), showing that high energy single electrons can be transmitted over a long distances.

Having demonstrated that our device can be used to produce a beam of ballistic electrons of tunable energy, we now scrutinize the distribution of the electron wavepackets in the time domain. This information is crucial for any scheme to synchronise the arrival of single-electrons in quantum circuits, but is generally difficult to measure because of insufficient experimental bandwidth.  Our method involves modulating the detector barrier voltage with a signal referenced directly to the pump control signal (see Fig.~1(a)), leading to a varying potential barrier height $V_b$.  As the electron energy is very well-defined, this gives a step-like modulation of barrier transparency (see Fig.~(a) and 2(b)).  Operating in this mode, we can extract information about the time distribution of the electron wavepackets in a straightforward way using our knowledge of the time-dependent barrier height.

\begin{figure}
\includegraphics[width=8.6cm]{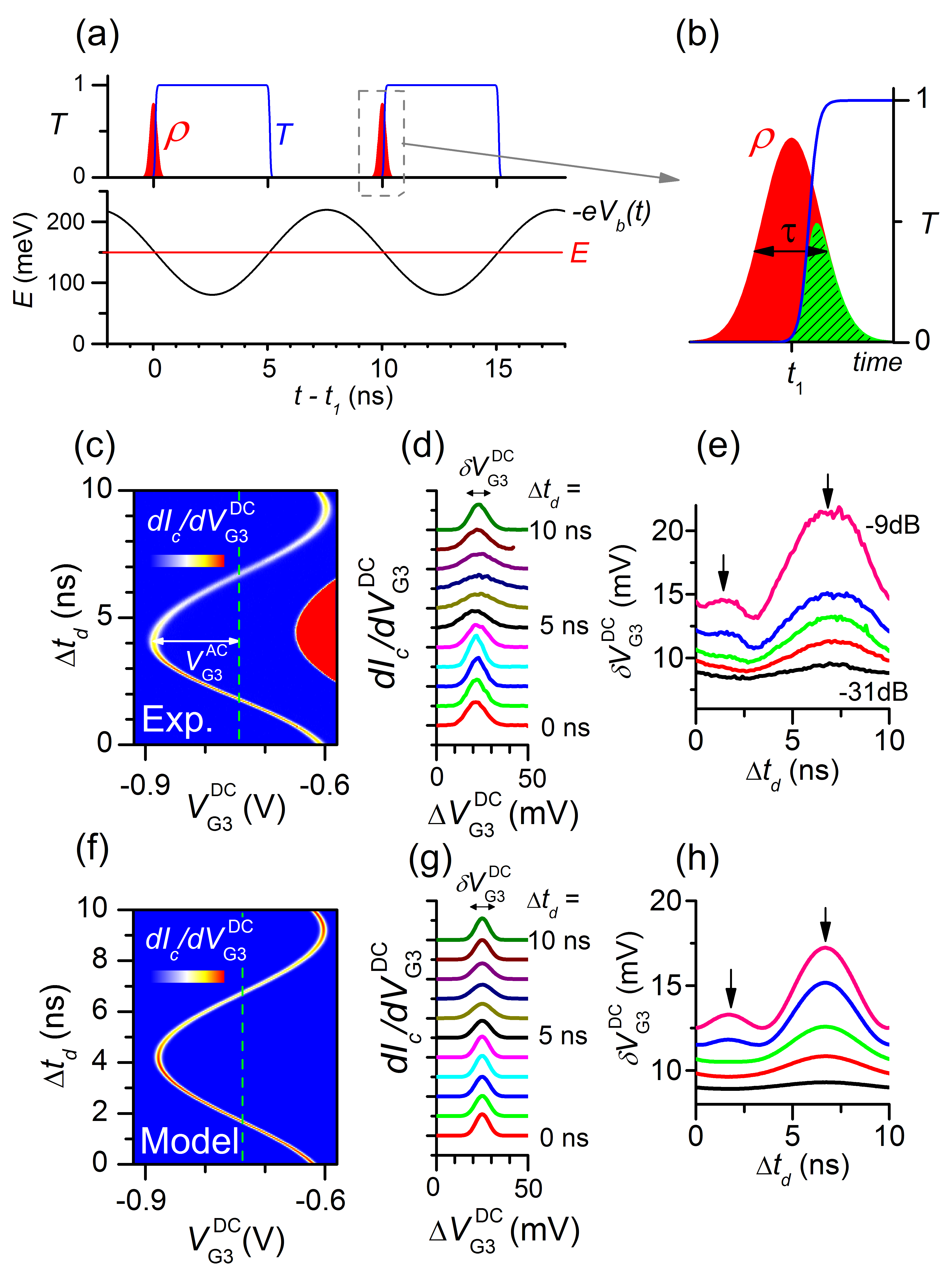}
\caption{
(a) Barrier transparency as a function of time for a modulated barrier height.
(b) Overlap of electron arrival probability distribution $\rho$ with barrier transparency $T$.
(c) Experimentally measured $dI_c/dV_{\rm G3}^{\rm DC}$ on a color scale as a function of $\Delta t_d$ and $V_{\rm G3}^{\rm DC}$ at $B = 12$~T and $A_d =$ -9~dB. The amplitude of the modulation of the threshold corresponds to the amplitude of the ac signal applied to the detector barrier, $V_{\rm G3}^{\rm AC}$. The solid red region on the right hand edge, while not important here, is due to the detector barrier becoming transparent for some fraction of the pump cycle. A dc current then flows, driven by a small bias voltage induced by the pump.
(d) Line cuts through the peaks in 2c showing extra broadening around $\Delta t_d = 7$~ns.%
(e) Full width at half maximum $\delta V_{\rm G3}^{\rm DC}$ of the peaks in 2c as a function of $\Delta t_d$ for different attenuation $A_d$ to the detector line (extracted by fitting to a Gaussian function). Data are offset for clarity. Two peak structure is indicated with two arrows.
(f) Numerical calculation using the model shown in 2a.
(g) Cut through model data equivalent to experimental data 2d.
(h) Calculated line width in the model using pickup and applied drive amplitudes $dV_{\rm CT}/dt$ = 40~mV/ns, $V_{\rm G3}^{\rm AC}$= 0-140~mV.
}
\label{f2}
\end{figure}

We apply a signal $V_{\rm G3}^{\rm AC}$ to the detector barrier gate at the same frequency as the pump modulation but via a time delay composed of fixed ($t_d$) and adjustable parts ($\Delta t_d$) and amplitude controlled by attenuation $A_d$ (see Fig.~1(a) for schematic). When this signal is applied, the blocking threshold in $V_{\rm G3}^{\rm DC}$ varies sinusoidally as $\Delta t_d$ is varied (see Fig.~2(c)). This suggests that the arriving electrons `sample' the instantaneous barrier height over a very short time duration compared to the pump cycle time (10~ns). The oscillating barrier voltage then acts only as a quasi-static shift in the barrier height, and a shift in $V_{\rm G3}^{\rm DC}$ is all that is required to restore the blocking effect. Closer examination reveals that the shape of the $V_{\rm G3}^{\rm DC}$ scan is also modified, with additional broadening appearing in the modulated barrier case. This is because the variations in barrier height are not negligible during the time taken for the electrons to traverse the barrier.

In the presence of a time varying barrier, the mean number of electrons transmitted is determined by an average of the barrier transmission $T(E,t)$ weighted by the distribution of electron arrival time and energy. We consider an electron probability density $\rho(E, t)$ peaked at time $t_1$ and $E_1$, but with a width in time of $\tau$ and of $\delta E$ in energy. The overlap of this with the transmission probability, dictated by the time varying barrier height, gives the current $I_c = ef\iint \rho T dt dE$. This situation is sketched in Fig.~2(b), where a cut in the time domain is shown for a fixed energy. When the leading/trailing edges of the single electron wavepacket enter the transmission window, as in the hashed region in Fig.~2(b), this gives $I_c \neq 0$ even for $V_{\rm G3}^{\rm DC}$ values far from the threshold for the static detector case. We can deduce the temporal width of the probability distribution $\tau$ from this broadening by rotating the phase of the barrier modulation. While the finite energy distribution also introduces broadening, it does not depend on this phase, and can be separated out.

\begin{figure}
\includegraphics[width=8.6cm]{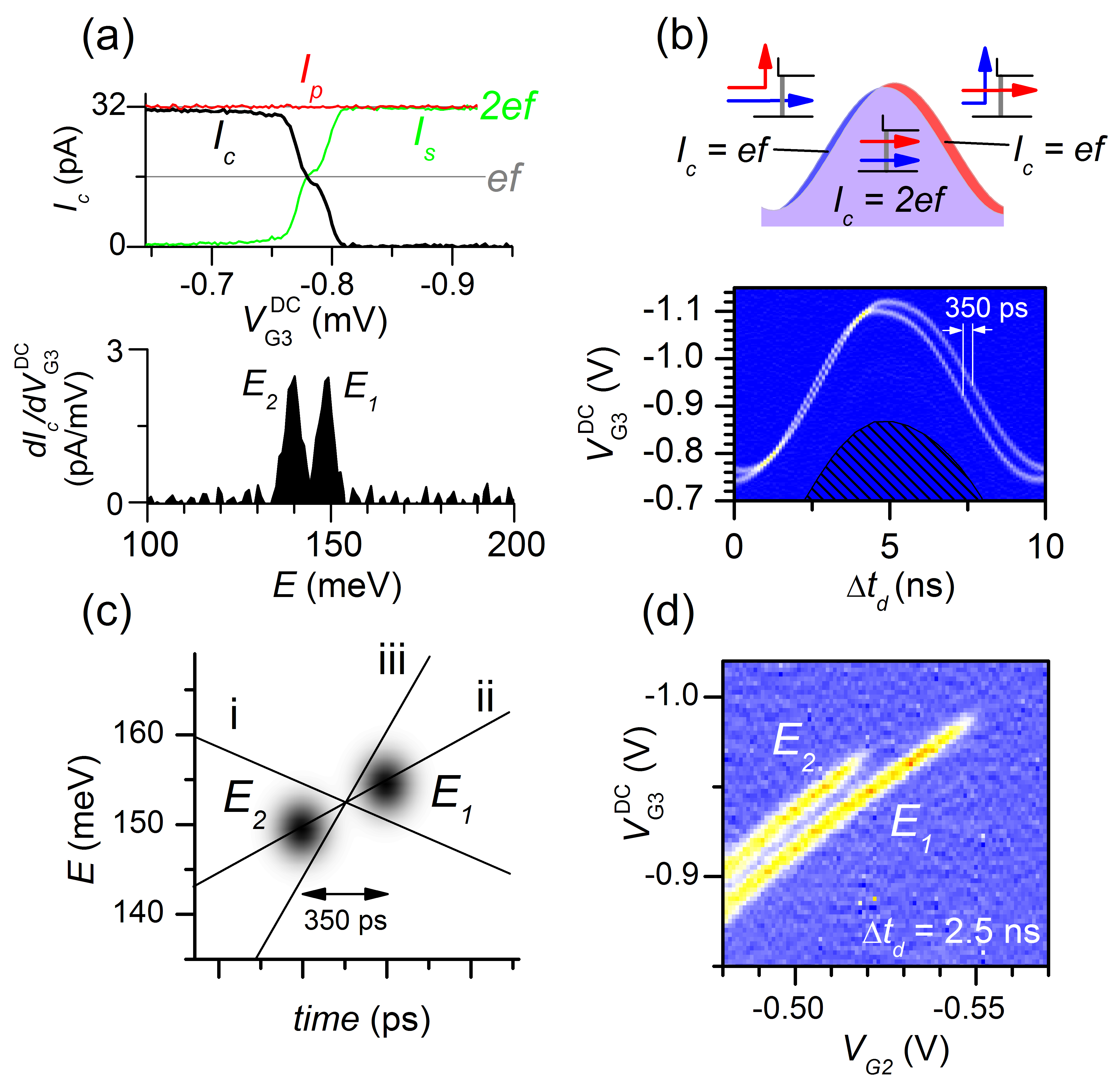}
\caption{
(a) Energy scan of the two-electron pump beam showing splitting of energy peaks. %
(b) Upper panel: Schematic of observed current while using modulated barrier on the two-electron beam. Lower panel shows experimentally observed $dI_c/dV_{\rm G3}^{\rm DC}$ on a color map. Vertical lines show the oscillation mid-point for the two lines separated by $\sim$ 350~ps.%
(c) Time domain picture for the filtering of the arrival of the two electrons (whose probability density is represented by a diffuse circle). When the barrier is static or moving slowly, low energy electrons are blocked (case $i$). For rapidly moving barriers (faster than the threshold case $ii$) the later, higher-energy electron can be blocked instead (case $iii$).%
(d) Spectral map in the crossover regime ($\Delta t_d = 2.5$~ns) showing the reversal of the order of the $E_1$ and $E_2$ lines.}
\label{f3}
\end{figure}

In Figs.~2(d,e) we show the experimentally observed broadening for different phase shift (controlled by $\Delta t_d$). In addition to a width of $\delta V_{\rm G3}^{\rm DC} \sim 8 $mV, which is attributed to the energy broadening due to the electron emission energy distribution and the finite broadening of barrier transmission threshold, we see a large enhancement in the width at $\Delta t_d \sim 7$~ns and a smaller enhancement at $\Delta t_d \sim 2$~ns. We note that these values of $\Delta t_d$ are where the applied modulation $V_{\rm G3}^{\rm AC}$ is changing most rapidly (compare with Fig.~2(c)), and hence where we expect the largest broadening due to the arrival time distribution. The disparity in height of these two peaks originates in the cross-talk signal from the oscillating voltage operating the pump; the large peak occurs when cross-talk and applied signals reinforce, while other peak is suppressed because of partial cancelation of the two signals.

The results of a numerical calculation, which includes a contribution from the cross-talk signal, are shown in Fig.~2(f-h). This captures the essential features of the experimental data and allows us to extract information about the electronic wavepacket size (see supplemental information for details). We conclude that the time of arrival distribution width is $\tau \sim 80$~ps, less than 1\% of the cycle time. This short distribution is due to a combination of the fast emission allowed by the strong variations in barrier height in our single-electron source and near-linear dispersion of Landau levels at high energy. In other cases, the emission distribution is likely to be even smaller than this, for example at higher operation frequencies the ejection process is necessarily faster and hence a shorter arrival time distribution is expected~\cite{Giblin2012}. This time-domain technique, which is effectively high bandwidth charge detection, can be used to measure these electron emission times or used to track the dispersion of single electron wavepackets over long distances.

The narrow emission energy range and short wavepacket time also allow us to explore new techniques for controlling electron beams consisting of multiple electrons. In our pump, when $V_{\rm G2}$ falls below a certain value, two electrons per cycle are loaded and ejected from the pump per cycle, changing the output current from $ef$ to $2ef$~\cite{Kashcheyevs2010}. The energy spectrum of the beam then features two separate peaks in the energy spectrum, as seen in Fig.~1(c) and in more detail in Fig.~3(a). In Fig.~1(c), one peak (labelled $E_1$) is perfectly continuous across the transition from one to two-electron pumping. The other (labelled $E_2$) abruptly ends when $V_{\rm G2}$ crosses the boundary to one-electron pumping. This second peak corresponds to the emission of one of the two electrons captured by the dot, leaving the other electron ($E_1$).

Using our modulated barrier technique we have found that this splitting of the energy spectrum is due to both differences in emission energy, and a relative delay in arrival of the two electrons at the detector. This means that for $V_{\rm G3}^{\rm DC}$ values between these peaks one of the emitted electrons crosses the detector and the other is reflected into the side channel.

Under strong modulation of the detector barrier, the $E_1$ and $E_2$ thresholds oscillate with $\Delta t_d$, as in the one-electron case (Fig.~3(b)). However, the presence of a shift on the $\Delta t_{d}$ axis of ~350~ps indicates that the electrons do not actually arrive at the detector at the same time, and hence encounter different barrier heights. The splitting in the measured spectrum does not depend solely on the emission energy, but also on delays in electron arrival. Figure~3(c) shows a model of the time and energy distribution of the two incoming electrons to illustrate this effect.

The order in which the electrons are blocked as the barrier is raised can be reversed depending on the time dependence of the barrier height. This is the origin of the crossing of the two threshold lines (Fig.~3(b)). In this regime ($1.5<\Delta t_d < 4$~ns) $I_c \simeq ef$ and $I_s \simeq ef$ (see Fig.~3(b)) but the spectral map (Fig.~3(d)) appears `flipped' with respect to that in Fig.~1(c) because the allocation of the first/second electron to collector/side channel has been reversed, in essence directing electrons according to order of arrival. This gives us a means of splitting a two-electron beam into two output channels in any combination we choose, with a known time delay between emission. This may be a useful technique for dividing electrons that are initially in a quantum superposition in the pump into two separate arms of a quantum device. The timing information is also critical for synchronising electron arrival, for instance in an electronic form of a Hong Ou Mandel interferometer~\cite{Giovannetti2006,Jonckheere2012}.

We have presented a way of controlling/conditioning single-electron wavepackets emission both in time and energy domains, and characterising their ballistic transport through a channel in the solid state.
Information that can be gleaned from this technique is not only the energy spectrum and scattering rate of the emitted electrons, but detailed time domain information that would be very difficult to obtain any other way. The ability to trap multiple electrons in a single confined system, then subsequently re-emit these on demand into multiple controlled output channels represents a possible new resource for the creation of experimental probes of entanglement in electronic systems.

We would like to acknowledge useful conversations with Stefan Ludwig and Heung-Sun Sim. This research was supported by the UK Department for Business, Innovation and Skills, NPL's Strategic Research Programme and the UK EPSRC.


\clearpage

\section{Supplemental Information}

{\bf Calibration of energy barrier:~}
To calibrate the energy barrier we followed a procedure similar to that used previously for continuous electron sources~\cite{SISchinner2009,SITaubert2011,SITaubert2011a}. In brief, we applied a source bias $V_{E}$ behind one of the electron pump gates, which is set to a voltage similar to that used in pump operation. The pump, side and collector terminals are connected together via ammeters, as in the single electron experiment. We then increase $V_{E}$ until this marginally exceeds the barrier threshold and a continuous current of hot electrons flows into the central channel. We assume that the electrons are essentially emitted with an energy corresponding to the source drain bias.  We then scan the energy detector barrier which, as in the single electron case, serves to block or permit the electron current into the collector $I_c$. Such data is shown in Fig.~S1. The horizontal dashed line indicates the minimum bias voltage required for current to be driven across the source barrier. The slight slope is because of a small capacitive coupling to the detector barrier gate. The tilted line represents the threshold to block the injected hot electrons. This edge is most sharply defined when the input current is small, near $V_E \sim -150$~mV boundary, and becomes more broad at larger $V_E$ where the current is larger (in the white region in the bottom right hand corner the current is out of range of this color scale. By finding the values of $V_{\rm G3}^{\rm DC}$ and $V_{E}$ at which the collector current is blocked, we identify that $dV_{\rm G3}^{\rm DC}/dV_{E} \simeq 3.0 \pm 0.2$ at the energies where we are operating. The features of our pump data which we ascribe to phonon emission (corresponding to an interval of $\sim$36~meV) are consistent with this relative calibration. To fix an absolute energy scale we consider that for $V_E = -147$~mV, the blocking threshold is $V_{\rm G3}^{\rm DC}=-0.79$~V. Near this point we use this as an approximate linear calibration $E = 0.147-(V_{\rm G3}^{\rm DC}+0.79)/3.0$ in units of eV. While this is satisfactory for the energy range of our electron pump emission, deviations from linearity occur at lower energy due to screening effects. For example, a certain voltage is required before the detector gate depletes the 2DEG enough to cut off transport at the Fermi energy $V_{\rm G3}^{\rm DC} = -0.2~$V. For measurements in this range a different calibration would be required.x

\begin{figure}
\includegraphics[width=8.5cm]{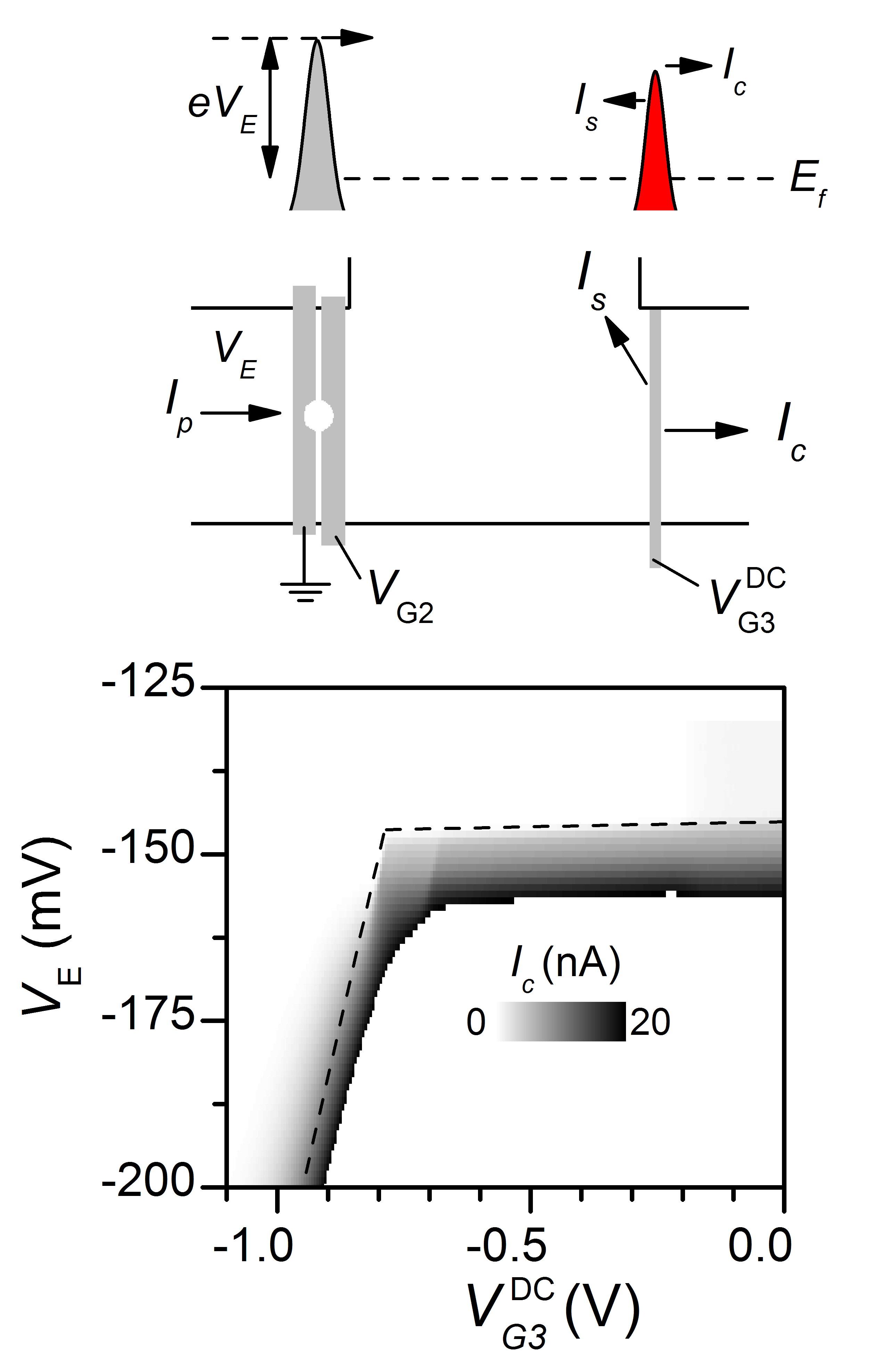}
\caption{Upper panel: Potential profile across device. A source bias is applied on the far side of one of the pump gates. Lower panel: Collector current on a grey-scale as function of bias voltage $V_E$ and detector barrier $V_{\rm G3}^{\rm DC}$. $V_{\rm G2}$ is set to a value of $\sim -0.4$~V and $V_{\rm G1}$ is grounded. The magnetic field is set to 12~T.)
}
\label{f4}
\end{figure}

{\bf Calculation of inelastic scattering length:~}
We estimate the scattering length by assuming that the fraction of pumped electrons $P$ in the main step in $I_c$ (rather than that in the phonon side peaks or any current detected in the side channel) are those that have not suffered any inelastic scattering in the distance $d \approx 3~\mu$m between source and detector. For $B = 12$~T, where the phonon side peaks are weak $P \gtrsim 0.95$. Assuming a constant scattering rate, giving an exponential scattering length, this gives $L_e = -d/\ln(P) \gtrsim 40~\mu$m.

{\bf Calculation of line broadening:~}

To calculate the collector current for a certain distribution of electron energy and arrival time we model a time and energy dependent transmission probability, convolved with a 2D gaussian distribution in electron energy and arrival time. The transmission function is based on a model detector response and consists of a broadened step-like transmission as a function of energy.
\begin{equation}
T_B = \left[\exp\left(-(eV_b+E)/\Delta_B\right)+1\right]^{-1}
\end{equation}
where $\Delta_B$ is the broadening. The threshold position is controlled by the relative barrier height, $-eV_b-E$, where $V_b$ is determined by the voltages applied to the detector barrier voltage, both continuous and oscillating.
The net barrier height is then given by
\begin{equation}
V_b =  \alpha \left[V_{\rm G3}^{\rm DC} + V_{\rm G3}^{\rm AC}\sin [\omega(t -t_d - \Delta t_d)] + V_{\rm CT}(t)\right]
\end{equation}
where $\alpha$ converts the values into potential at the 2DEG, rather than gate voltage. The variable delay $\Delta t_d$ is controlled by the programmable delay line. The fixed (but unknown) effective delay $t_d$ is due to differences in cable length and other differences in the signal paths of the two lines. For the calculations this is set to reproduce the location of the origin of $\Delta t_d$ (this simply shifts the origin). The pick up signal $V_{\rm CT}$ originates in cross talk from the pump (see below).

The average current for these conditions was calculated by sampling $T_B (E,t)$ with a normalised Gaussian kernel
\begin{equation}
\rho(E,t) = \frac{1}{2\pi\sigma_t \sigma_E}e^{-(t-t_1)^2/2\sigma_t^2-(E-E_1)^2/2\sigma_E^2}
\end{equation}
numerically on a finite energy and time mesh indexed by $i,j$,giving
\begin{equation} I_{c} = ef \sum_{i,j} T_{i,j}K_{i,j}.\end{equation}
As we are modeling data where the pump source parameters are fixed, the peak energy, $E_1$, and time $t_1$ are fixed, along with the widths of the distribution $\sigma_t$ and $\sigma_E$. The transmission probability is controlled via the detector parameters, $V_{\rm G3}^{\rm DC}$, $\Delta t_d$ and $V_{\rm G3}^{\rm AC}$.

Experimental values of the energy threshold  $V_{\rm G3}^{\rm DC} \sim -0.74$~V and modulation up to $V_{\rm G3}^{\rm AC} = \pm 0.15$~V are used (the latter can be directly measured e.g. from Fig. 2c). The baseline level of broadening due to the finite electron energy distribution and the barrier function (these individual contributions are difficult to separate) is $\sim 8~$~mV with a static barrier. Peaks in additional broadening occur when the detector changes height most rapidly, in either direction, but size of the effect will be asymmetric, as observed experimentally, in the presence of any other time dependent signals that partly cancel or reinforce this. From the experimental data we find that $dV_{\rm CT}/dt \sim 40$mV/ns at $t = t_1$ would explain the mismatch, and allow the estimation of the time width of $\tau \sim 80~$ps as in the main text.

We have detected a clear cross-talk signal from the pump in separate experiments, namely measuring the detector conductance while modulating the pump barrier (with the electron pump configured to a non-pumping regime). The value of $V_{\rm G3}^{\rm DC}$ required to block conduction at the Fermi energy shifts to more negative voltage due to a reduced barrier height, induced due to the coupling of the pump barrier to the detector gate and the surrounding 2DEG. This is almost identical to the shifting up and down of the single-electron threshold (see Fig. 1b) and is visible in the shape of the off-scale DC current in Fig. 2c and Fig. 3b.

{\bf Contributions to line-broadening:~}
In addition to the contribution from the finite electron arrival time distribution, we identify and quantify several other processes which could potentially give a broadening of the electronic distribution.
\textit{Quantum broadening:} Firstly, the minimum emission broadening is dictated by the finite coupling energy during electron emission, corresponding to an energy spread of $0.01~$meV for a 100~ps emission window.
\textit{Changes in pump energy level during emission:}
Previous calculations \cite{SIKaestner2008,SIFletcher2012} indicate that the energy level in the dot is not strictly constant during the pumping process, potentially leading to a smearing of the energy output. This would not have any particular dependence on $\Delta t_d$, but would give an approximately constant contribution to the broadening.
We can estimate the contribution from the size of the rf voltage applied to the primary pumping gate, which is by far the fastest changing scale in the problem. The power applied to this gate corresponds to a voltage excursion of approximately $\pm 0.36~V$ around the constant value $V_{\rm G1} \sim -0.4$~V. The maximum variation in the voltage applied to this gate is then $\sim 2 \times 10^8$~V/s. To estimate an energy distribution we have to estimate likely changes in dot energy from this change in gate voltage considering several factors. Firstly, the actual barrier potential will be smaller by a factor similar to that found in the detector calibration described above. Secondly, this barrier height only changes the dot energy level indirectly; this is not a direct `plunger' gate. This is particularly true as the barrier becomes very tall near the point of particle ejection. Finally, the entrance barrier potential is near its most negative value when electrons are emitted, so the rate of change of this barrier height is slowed near the moment of particle ejection. We estimate that these factors give an energy spread attributable to the pump modulation of $<1$~mV, much smaller than the values seen experimentally.

\textit{Back action of the energy barrier on the pump: }We have found that when strongly modulating the detector barrier this has a small but detectable effect on the operating point of the electron pump. The main observation is that the boundaries of the pump current plateaux as a function of $V_{\rm G2}$ move by up to $\pm5$~mV, varying sinusoidally as the detector modulation delay $t_d$ is varied. This is a small shift compared to the width of the plateaux ($\sim$30-40~mV), and is only visible when the detector barrier is being modulated most strongly.
Using the observed linear relationship between the peak position in $V_{\rm G2}$ and $V_{\rm G3}^{\rm DC}$ (Fig. 1c) this can be translated into a variation in emission energy of up to $\sim 3$~meV for the largest detector modulations that we have applied. This effect on the source energy is essentially disguised by the much bigger effect (15 times bigger) that modulation has on the detector threshold (Fig. 2c). We can estimate how much the time dependence of this effect might give an apparent energy broadening. At the value of $\Delta t_d$ that would create the fastest perturbation to the emission energy, and using our arrival time distribution time as an upper limit on the emission time, we expect a line broadening of 0.6~mV. This is about ten times smaller than the overall values seen in Fig. 2d, but might plausibly distort certain features of the data when operating at high modulation, for instance, introducing a weak ($\sim$10\%) sinusoidal modulation to the broadening, on top of the double peak structure. While the overall structure of the data are dominated by the time broadening effect, this introduces an uncertainty of $\sim 20\%$ into the exact value of the arrival time distribution. This is comparable with other contributions, like the detailed choice of the shape of the barrier function and assumptions about the arrival time distribution.

\end{document}